\documentclass[journal=jacsat,manuscript=article]{achemso}

\usepackage[version=3]{mhchem} 

\usepackage{hyperref}

\author{Gyehyun Park}
\author{YounJoon Jung}
\email{yjjung@snu.ac.kr}
\affiliation[Unknown University]
{Department of Chemistry, Seoul National University, Seoul 08826, Korea.}

\title[An \textsf{achemso} demo]
  {Many-Chain Effects on the Co-nonsolvency of Polymer Brushes in a Good Solvent Mixture}

\abbreviations{IR,NMR,UV}
\keywords{American Chemical Society, \LaTeX}

\begin{document}







\begin{abstract}
Polymer brushes normally swell in a good solvent and collapse in a poor solvent. An abnormal response of polymer brushes, so-called co-nonsolvency, is the phenomenon where the brush counter-intuitively collapses in a good solvent mixture. In this work, we employed molecular dynamics simulations to investigate the structural properties of the grafted polymers on the occurrence of co-nonsolvency. Brushes with various grafting densities were considered to study the effect of topological excluded volumes on the co-nonsolvency. We found that the brush height follows a novel scaling behavior  of the grafting density $h \sim \sigma_{\text g}^{0.71}$ in the co-nonsolvent mixture. Using the scaling exponent and Alexander-de Gennes theory, an analytic function that predicts the monomer density was obtained. The many-chain effects in the co-nonsolvent lead to the formation of both intermolecular and intramolecular bridging structures. Increasing the grafting density entails lower looping events occuring because of  the intermolcular bridging, causing diverse structural properties.  We report how the average thickness, the polymer orientation, and the looping probability varies as the grafting density increases. Based on these observations, we constructed a phase diagram of the polymer brush system using the average thickness and orientation as order parameters. Our simulations and analytical results reveal the nature of co-nonsolvency in polymer brushes in an explicit way and will help to provide practical guidelines for applications such as drug delivery and sensor devices.
\end{abstract}

\section{Introduction}

Polymer brushes are a collection of polymer chains that are tethered to a metal surface or a biological membrane. These bundles of grafted polymers change structure on the application of external stimuli such as temperature, light, pH, and solvent. Because of their stimuli-responsive nature, polymer brushes have been utilized in various biological systems and applied to develop synthetic materials with sensory functions.~\cite{Howarter07,Stuart10,Lee10,Chen10} Typical applications include smart drug eluting devices that can effectively regulate pharmacological activity in the human body.~\cite{Ruiz08,Yu13,Alvarez-Lorenzo14} Stimuli-responsive polymer brushes with tethered biocides or drugs can undergo a directed conformational transition when an external stimuli is applied at the target site. The transition between the swollen and collapsed states of the brush induces a change in the biocide exposure and the diffusion of drugs such as vancomycin, allowing the modulation of pharmacological action. 

The collapse of a polymer in a mixture of good solvents, dubbed co-nonsolvency, has attracted a great attention as an unexpected response to the solvent stimulation of polymer brushes. ~\cite{Sui11, Yu13, Yu16, Sommer17, Chen17} In polymer physics, polymers expand in a good solvent and collapse in a poor solvent. However, in the co-nonsolvency phenomenon, polymer brushes contract in a mixture of good solvents even though the brush expands in each of the good solvents alone. Co-nonsolvency is not only an intriguing theoretical issue to explore from the viewpoint of fundamental polymer physics but also important novel stimuli-responsive effect with practical applications. Accordingly, many experimental,~\cite{Hao10, Sui11, Yu13, Yu16, Mukherji16} computational,~\cite{Walter12, Mukherji13, Mukherji14, Mukherji15, Mukherji16,Dalgicdir17} and theoretical studies~\cite{Tanaka08, Mukherji14, Budkov14, Budkov16, Chen17, Budkov17, Sommer17, Budkov18} have been performed in recent decades. Experimentally, co-nonsolvency has been observed in various combinations of polymers and solvent species and is not limited to specific systems.~\cite{Hao10, Sui11, Winnik92, Winnik93, Mahdavi09, Pagonis04, Orakdogen06, Saunders97, Crowther98, Mukherji16, Walter12, Yu16} Among those sets of polymer-solvent mixtures widely used to study co-nonsolvency systems are poly({\it N}-isopropylacrylamide) (PNIPAM) and methanol/water mixture. The abrupt collapse of PNIPAM brush occurs when a small fraction of methanol ($\sim 20\text{ mol}\%$) is added to water, resulting in changes in the thickness and the energy dissipation, as shown by atomic force microscopy (AFM)~\cite{Sui11} and quartz crystal microbalance (QCM)~\cite{Liu05} experiments. However, when the methanol fraction is increased to more than $40\text{ mol}\%$, the brushes gradually re-swell and regain their pure-water conformation. This asymmetric change with solvent composition has also been observed with respect to the dynamic,~\cite{Hao10,Huan17} synthetic~\cite{Pomorska16}, and mechanical properties,~\cite{Qi14,Yu16,Yu17} extending the notion of the co-nonsolvency beyond the structural response.

To explore the co-nonsolvency theoretically, Mukherji, Kremer, and Marques proposed the single polymer adsorption (SPA) model which assumes that all sites capable of contacting with solvent molecules in the polymer chain can act as adsorption sites.~\cite{Mukherji14} In this model, the better solvent occupies two adsorption sites simultaneously, having a stronger affinity to polymer than the good solvent, and a loop is formed in which monomers are bridged through the better solvent. The SPA model considers thermodynamics in the formation of the loop structures produced by the intramolecular bridging structures,~\cite{Jacques90} and co-nonsolvency is explained as the preferential solvation of the better solvent to compensate for the entropy loss incurred from demixing and looping. Nevertheless, the polymeric system that exhibits co-nonsolvency contain many polymer chains, except for a very few cases such as single chain experiment.~\cite{Zhang01} Systems such as micro-gels or polymer brushes bounded by interchain interactions cannot be described by the SPA model. To overcome this limitation, the SPA model was later extended by Sommer to account for the co-nonsolvency of the polymer brushes.~\cite{Sommer17} This model, called the adsorption-attraction (AA) model, combines the concepts of the SPA model and the Alexander-de Gennes brush theory.~\cite{Alexander77, deGennes80}. \color{blue}Galuschko and Sommer demonstrated that the AA model well-describes the continuity of the collapse transition in the co-nonsolvency response compared with the explicit model.~\cite{sommer19} However, the AA model assumes that the elastic free energy of chains using mean-field approach and does not account for the fluctuation between looping and unlooping conformation free energies. As a result of the lack of thermodynamic degrees of freedom that distinguish between intermolecular and intramolecular bridging free energies, this model is limited in explaining the structural properties such as the polymer orientation and looping probability, which originate from the variation of the inter or intrabridging conformations.

Because of the model limitations and the lack of theoretical studies about polymer brushes in co-nonsolvent mixtures, the fractal nature of polymer brushes in co-nonsolvent mixture has yet to be found so far. In a standard good solvent, the height of overlapping brush scales with $h \sim N {\sigma_{\text g}}^{1/3}$ for a polymer brush with the grafting density $\sigma_{\text g}$ and polymers of $N$ monomeric units~\cite{Alexander77,deGennes80} and in the poor solvent, , the height scales with $h \sim N {\sigma_{\text g}}^{0.8}$.\cite{Moh11} The Flory exponent $\nu$, which determines the size of "correlation blob" in the overlapping brush of Alexander-de Gennes brush, is known to be $\nu=0.588$ in the good solvent and $\nu=1/3$ in the poor solvent.~\cite{Rubinstein03} Finally, the looping exponents, which estimate the tendency for monomers to loop back to the origin, are known to be $\alpha=2.2-2.4$  in a good solvent\cite{Podtelezhnikov97,Debnath04,Thirumalai99,Toan08} and $\alpha=1.0$ or $1.6-1.7$ in a poor solvent.\cite{Toan08,Debnath04}  Although it has been observed that, phenomenologically, polymer brushes in co-nonsolvent mixtures behave as though they were in a poor solvent, there is no guarantee that the scaling behavior in the co-nonsolvent mixture will be equal to that in the poor solvent.

In this work, we report the scaling behavior of the brush height in co-nonsolvent mixture with respect to the grafting density via molecular dynamics simulation. On the basis of this scaling exponent, we evaluated the Flory exponent $\nu$ in the co-nonsolvent mixture using the Alexander-de Gennes theory. Finally, we obtained an analytic master curve that can be used to predict the monomer density profile of the maximally collapsed state in co-nonsolvent mixture. We also investigated the structural properties such as polymer orientation, looping probability and the radius of gyration explicitly. From these properties, we suggest a microscopic for the explanation for the experimentally observed phenomena of the co-nonsolvency brushes.

The outline of this paper is as follow. First, we describe the coarse-grained model and the molecular dynamics simulation method. Then, we present the structural properties of polymer brushes in a standard good solvent and compare the numerical density profiles with those obtained from the previous analytic curve. Subsequently, we discuss the scaling behavior of the brush height in a co-nonsolvent mixture and derive a novel master curve using the Alexander-de Gennes brush theory. Finally, we analyze the brush structure in the co-nonsolvent mixture using macroscopic properties, such as the average thickness, and the microscopic properties, such as bond orientation or looping probability. The phase diagram of the structural properties of the brush is also presented. We conclude with a discussion of the significance of our research and suggest important topics for future study. For additional details, electronic supporting information${\dag}$ (ESI) is provided. 

\color{black}
\section{Methods}
In this work, we use the bead-spring polymer model developed by Kremer and Grest~\cite{Kremer90} to perform molecular dynamics simulations to study the co-nonsolvency phenomenon. Although the model is not atomically resolved, it was chosen because many the features observed in polymer brush systems are quite generic.~\cite{Mukherji14, Hao10, Sui11, Winnik92, Winnik93, Mahdavi09, Pagonis04, Orakdogen06, Saunders97, Crowther98}
The polymer had a degree of polymerization $N=50$  and were grafted to a substrate, the so-called wall. Each monomer (m) in the polymer is connected by a finitely extensible nonlinear elastic (FENE) and Weeks-Chandler-Andersen (WCA) potential,
\begin{equation}
U_{\text{FENE}} =
\begin{cases}
-\frac{1}{2}k{R_{0}}^{2}\text{ln} \bigg[ 1-{\Big( \frac{r}{R_{0}} \Big)}^{2} \bigg],     & \quad r<R_{0}\\
\infty,  & \quad r\geq R_{0}\\
  \end{cases} \label{eq:U_FENE}
\end{equation}
\begin{equation}
U_{\text{WCA}} =
\begin{cases}
4\epsilon_{\text{mm}} \bigg[ {\Big( \frac{\sigma_{\text{mm}}}{r}\Big)}^{12} 
- {\Big( \frac{\sigma_{\text{mm}}}{r} \Big)}^{6}+\frac{1}{4}\bigg],     & \quad \text{for } r \leq 2^{1/6}\sigma \\
0,  & \quad \text{elsewhere}\\
  \end{cases}
\end{equation}
having a finite extensibility $R_{0}= 1.5\sigma$ for the FENE potential, and a bond strength $k= 30\epsilon/\sigma^{2}$, $\sigma_{\text{mm}}=1.0\sigma$ and  $\epsilon_{\text{mm}}=1.0\epsilon$ for the WCA potential. The polymers are solvated in a mixture of two types of Lennard-Jones (LJ) particles: the so-called good and better solvent molecules. The LJ potential was applied to all possible pair interactions of monomers and solvent molecules. The full LJ potential was used for the interaction between a monomer (m) and a better solvent molecule (b),
\begin{equation}
U_{\text{LJ}} =
\begin{cases}
4\epsilon_{\text{mb}} \bigg[ {\Big( \frac{\sigma_{\text{mb}}}{r} \Big)}^{12} - 
{\Big( \frac{\sigma_{\text{mb}}}{r} \Big)}^{6}
-{\Big( \frac{\sigma_{\text{mb}}}{r_{\text c}} \Big)}^{12} 
+ {\Big( \frac{\sigma_{\text{mb}}}{r_{\text c}} \Big)}^{6}\bigg],&\text{for } r \leq 2.5\sigma \\
0, &\text{elsewhere}\\
  \end{cases} \label{eq:U_LJ}
\end{equation}
where $\sigma_{\text{mb}}=1.0\sigma$ and  $\epsilon_{\text{mb}}=1.5\epsilon$. The WCA potential was used for the pair-wise interactions between the other types of species, 
\begin{equation}
U_{\text{WCA}} =
\begin{cases}
4\epsilon_{\text{ij}} \bigg[ {\Big( \frac{\sigma_{\text{ij}}}{r}\Big)}^{12} 
- {\Big( \frac{\sigma_{\text{ij}}}{r} \Big)}^{6}+\frac{1}{4}\bigg],     & \quad \text{for } r \leq 2^{1/6}\sigma \\
0,  & \quad \text{elsewhere}\\
  \end{cases}
\end{equation}
where the indices ij refer to different types of pairwise interactions between monomers, good solvent molecules (g), better solvent molecules, and wall atoms (w), except for the monomer-better solvent molecule interaction. The parameters were set to $\sigma_{\text{ij}}=1.0\sigma$ and  $\epsilon_{\text{ij}}=1.0\epsilon$.

Two explicit walls perpendicular to the $z$-axis were placed at the bottom ($z=0$) and at the top ($z=L_{z}$) of the system. The grafting wall was modeled as static particles composed of wall atoms and the first monomer of each chain. The velocities of these particles were set to zero during simulation to achieve stationary particles. The lattice spacing was set to $a=1.25 \sigma$ so that the solvent molecules were prevented from penetrating the substrates but, at the same time, the repulsion between adjacent chains was not too large.~\cite{Dimitrov07} Periodic boundary conditions in the $x$ and $y$ directions, which are parallel to grafting surface, were imposed. 

\begin{figure}[t]
\centering
 \includegraphics[height=3.5cm]{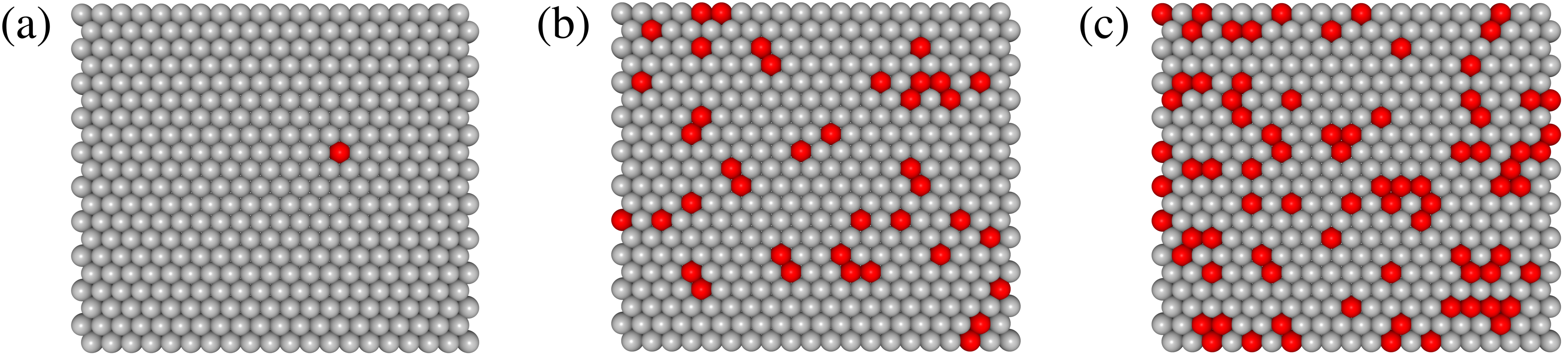}
  \label{P80}
  \caption{Simulation snapshots showing the bottom view of the grafting surfaces: (a) $\sigma_{\text g}=0.002$ (b) $\sigma_{\text g}=0.074$, and (c) $\sigma_{\text g}=0.148$. The gray and red spheres represent the wall atoms and the grafted monomers, respectively, which correspond to the first monomers in the polymer chain. The grafting sites were selected randomly in each independent trajectory.}
  \label{surface}
\end{figure}

\begin{figure}[t]
\centering
  \includegraphics[height=5.5cm]{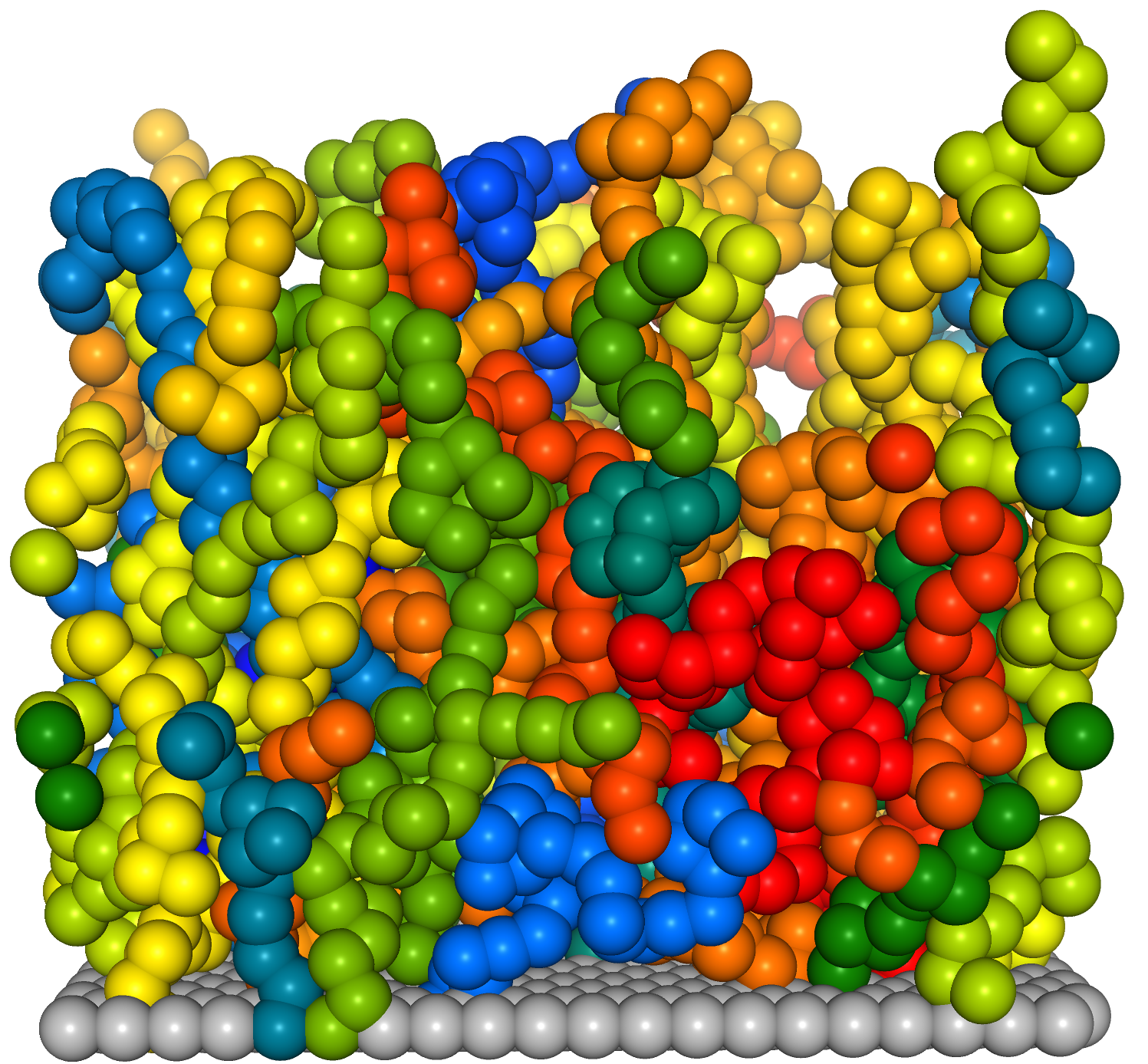}
  \label{high_T}
  \caption{Initial configuration of grafted polymers with random conformations obtained by the high temperature sampling ($T=1.5\epsilon/k_{\text{B}}$) for $M=40$ (corresponding to $\sigma_{\text g}=0.074$).}
  \label{ini_conf_and_high_T}
\end{figure}

Brushes corresponding to eight different surface coverages were prepared by varying the number of grafted polymers $M$, and the horizontal lengths of the system box $L_x$ and $L_y$, as listed in Table S1 in the ESI${\dag}$. The grafting density can be expressed as $\sigma_{\text{g}}=M/(L_{x}L_{y})$. Three representative brushes with $\sigma_{\text{g}}=0.002, 0.074$ and $0.148$ are shown in Fig.~\ref{surface}. For all the systems, the total number of solvent molecules is fixed by $N_{\text{tot}}=12000$. We simulated the grafted polymer with various solvent compositions. The mole fraction of the better solvent $x_{\text{b}}=N_{\text{b}}/N_{\text{tot}}$ was varied in intervals of $\Delta x_{\text{b}}=0.05$  in the the range $x_{\text{b}} \leq 0.2$ and in intervals of  $\Delta x_{\text{b}}=0.1$ in the range $x_{\text{b}} > 0.2$. We maintained the solvent density constant at $\rho=0.5\sigma^{-3}$ for all the systems, by varying the height of the system box $L_{z}$ depending on the number of grafted polymers. 

The $NVT$ simulation were performed using the velocity Verlet algorithm with an integration time $\delta t=0.005\tau$ where the unit time was scaled by $\tau=(\epsilon/m\sigma^2)^{1/2}$. The temperature was set to $T=0.55\epsilon/k_{\text{B}}$ using the Nos\'e-Hoover thermostat. To prepare the initial configurations of the grafted polymers with random conformations, we first designed polymers rod conformations in an array geometry. Subsequently, we sampled the grafted polymers at high temperature $T=1.5\epsilon/k_{\text{B}}$ without a solvent (Fig.~\ref{ini_conf_and_high_T}). The equilibration was carried out for $10^{5}\tau$. All simulations were performed using the GROMACS 4.6.7 simulation packages.~\cite{Pronk13}

\begin{table}[t]
\small
  \caption{\ Structural characteristics of the brushes with three grafting densities in the good solvent conditions : the average distance between grafting sites $\langle D \rangle$, the parallel and perpendicular components of the radius of gyration $R_{\text{g},xy}$ and  $R_{\text{g},z}$, and the maximum height of the brush $h$.}
  \label{table_brush}
  \begin{tabular*}{0.48\textwidth}{@{\extracolsep{\fill}}lllll}
    \hline
    $\sigma_{\text g}$ & $\langle D \rangle$ & $R_{\text{g},xy}$ & $R_{\text{g},z}$ & $h$ \\
    \hline
    0.002   & 22.36 & 3.55 & 2.54 & 18.42 \\
    0.074   & 3.68  & 2.94 & 3.90 & 25.53 \\
    0.148   & 2.60  & 2.64 & 4.91 & 28.70 \\
    \hline
  \end{tabular*}
\end{table}

\section{Results}
\subsection{Topological effects of brushes in a good solvent}

Different surface coverages of brushes were first characterized in a pure good solvent before studying the structures in the solvent mixture. As shown in Table~\ref{table_brush}, the in-plane component of the radius of gyration $R_{\text{g},xy}$ at the lowest grafting density ($\sigma_{\text{g}}=0.002$) is less than the average distance between the grafting sites of neighboring chains, $\langle D \rangle$=$\sigma_{\text{g}}^{-1/2}$. In this regime, a chain can only avoid itself and does not interact with any other chains. However, as the grafting density increases, the similar values of $R_{\text{g},xy}$ and $\langle D \rangle$ were observed, which indicates that the overlapping of chains.

\begin{figure}[t]
\centering
  \includegraphics[width=8.7cm]{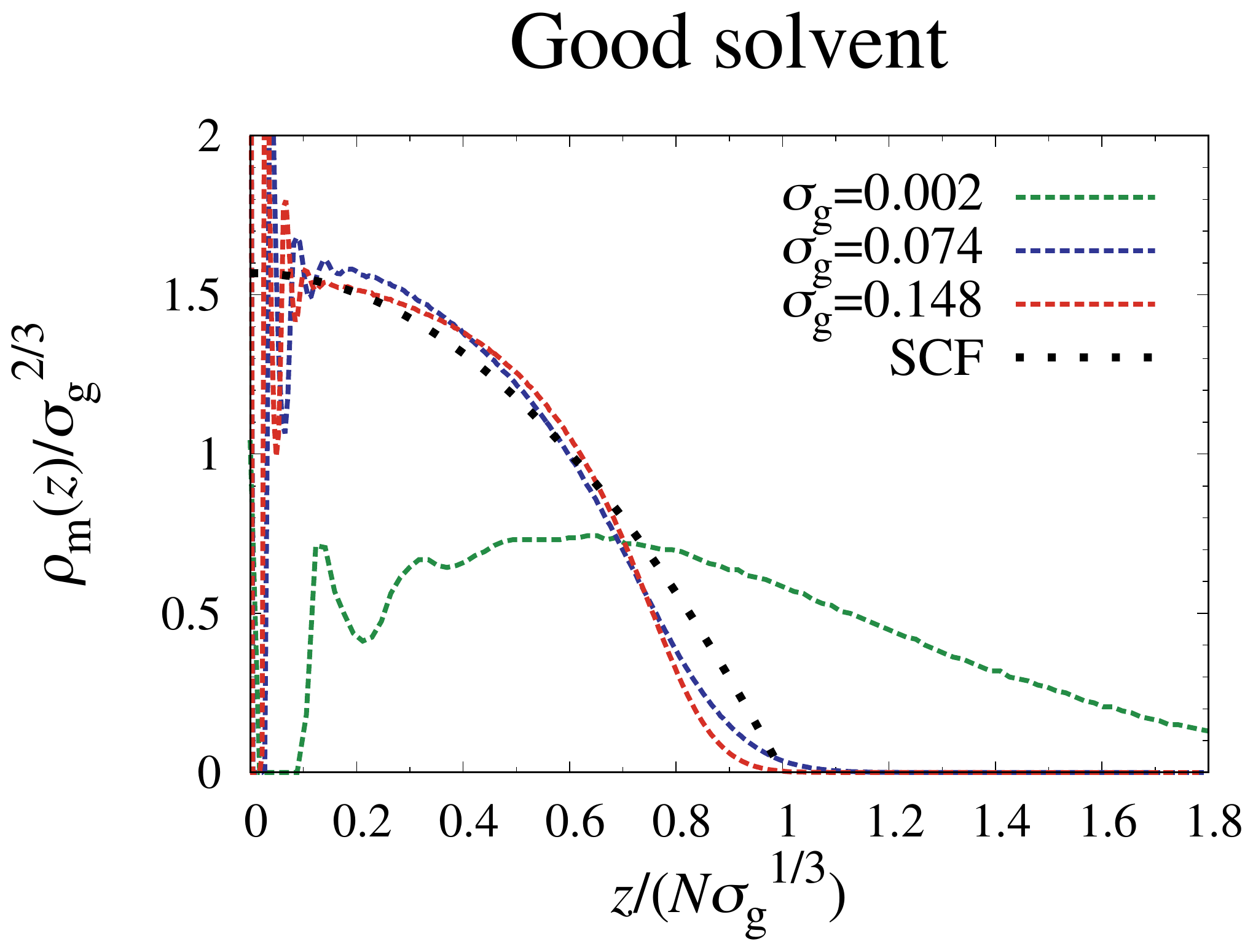}
  \caption{Monomer density scaled by ${\sigma_{\text g}}^{2/3}$ as a function of the vertical coordinate scaled by $N{\sigma_{\text g}}^{1/3}$ for comparison with the universal profile of the overlapping brush which was theoretically predicted by Milner.~\cite{Milner88} The black curve represents the SCF prediction given in Eq.~(\ref{eq:rho_m}).}
  \label{theory}
\end{figure}
 
To determine the threshold of the grafting density at which interchain interactions occur, we compared the monomer densities for three grafting brushes with those predicted by the self-consistent field (SCF) method\cite{Milner88} for the moderate and high surface coverages, as shown in Fig.~\ref{theory}. The rescaled monomer density profile at $\sigma_{\text g}=0.074$ and $0.148$, where weak interactions between the grafted polymers exist, shows parabolic decay consistent with the analytic expression in Eq. ~\ref{eq:rho_m}\cite{Milner88, He07} 
\begin{equation}
\frac{\rho_{\text m}(z)}{\sigma_{\text g}^{2/3}}=\frac{\pi^2}{8w}(H^{2}-Z^2)\Theta(H-Z) \label{eq:rho_m},
\end{equation}
In this equation, $H=h/(N{\sigma_{\text g}}^{1/3})$ denotes the rescaled brush height and $Z=z/(N{\sigma_{\text g}}^{1/3})$ is the rescaled $z$ coordinate, and $w$ is the strength of the excluded volume interaction between chains, respectively. $\Theta$ is the Heaviside step function. 
The analytic density curve in Eq.~(\ref{eq:rho_m}) is also consistent with the overlapping density curves reported from previous simulations.~\cite{He07, Chakrabarti90, Dimitrov07} 
In contrast to the intermediate and highest grafting density cases, the lowest grafting density case does not fall onto an analytical, master curve, which suggests that the characteristic effective potential between the chains at a lowest grafting density ($\sigma_{\text{g}}=0.002$) is clearly different from those at intermediate ($\sigma_{\text{g}}=0.074$) and highest ($\sigma_{\text{g}}=0.148$) grafting densities. Therefore, we carefully designed a mushroom brush regime ($\sigma_{\text{g}}=0.002$) to exclude the inter-chain interactions and overlapping brush regime ($\sigma_{\text{g}}=0.074$ and $0.148$) to include the inter-chain interactions. In particular, the intermediate grafting case results in a much longer correlation length $\xi$ than the high grafting case, which can be seen as the scaling relationship between the correlation length $\xi$ and the grafting density $\sigma_g$, where $\xi \approx \sigma^{-1/2}_{g}$. This indicates that more inter-chain interactions are present in the moderate grafting density brush than in the highest grafting density brush.

\subsection{Swelling-collapse-swelling transition}

\begin{figure}[t]
\centering
 \includegraphics[width=7.5cm]{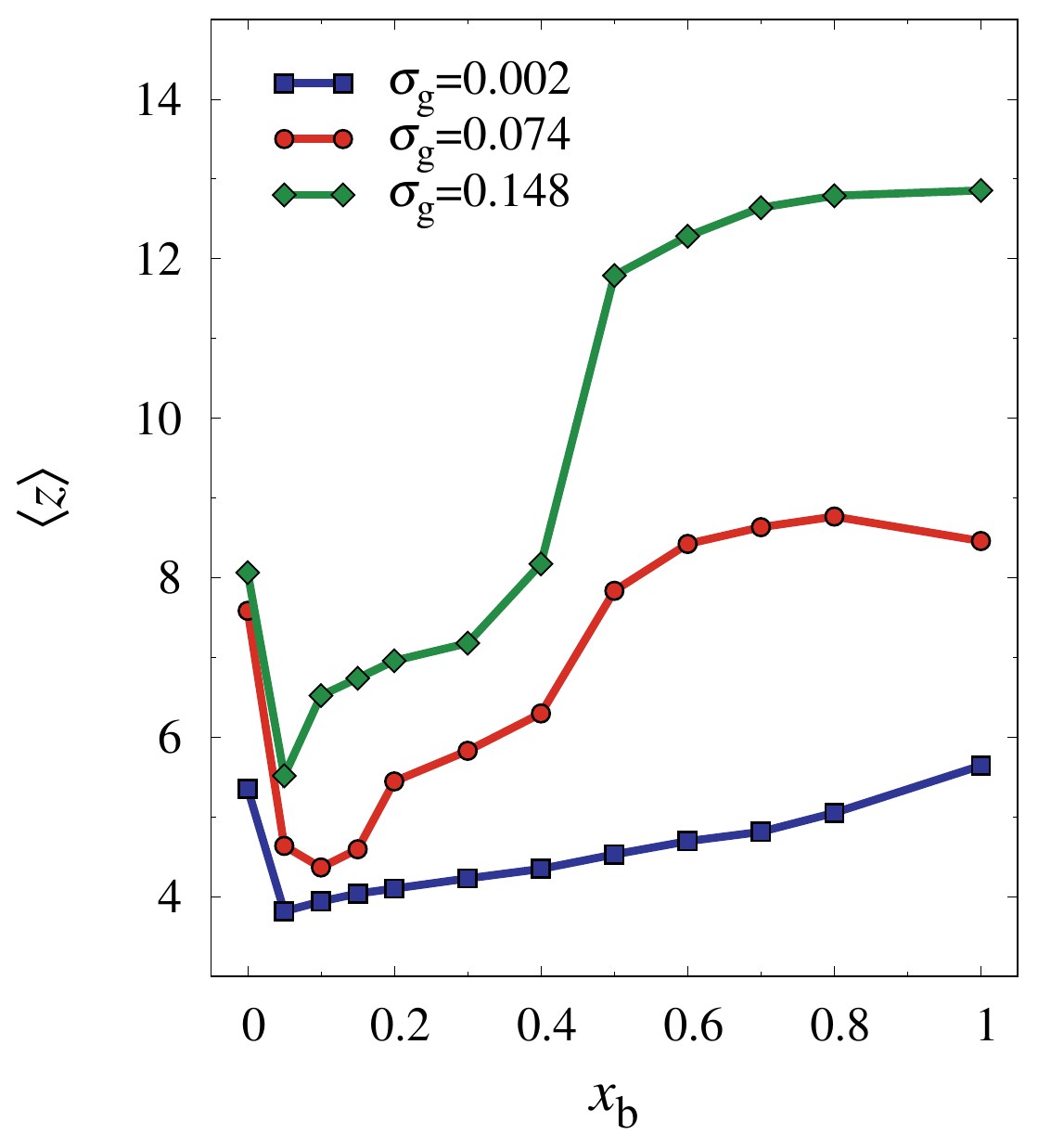}
\caption{Average thickness as a function of the better solvent fraction for three grafting densities. The values were obtained by time-averaging. The maximally collapsed co-nonsolvency window ($0.0<x_{\text b}<0.2$) was carefully observed in intervals of $x_{\text b} = 0.05$. The different point shapes denote the different grafting densities. The lines connecting the points having the same grafting condition shows different collapse-reentrant behavior in the brushes. }
\label{z}
\end{figure}

To investigate the co-nonsolvency behavior for the three different grafting brushes, we calculated the average thickness of the brush defined by
\begin{equation}
\langle z \rangle=\frac{\int_0^\infty z\rho_{\text m}(z)dz}{\int_0^\infty \rho_{ \text m}(z)dz}\label{eq:z}.
\end{equation}
Here, $\rho_{\text{m}}(z)$ denotes the density of the monomer at the height $z$ from the grafting surface. With the addition of more better solvent molecules until $x_{\text{b}}=0.05$, the brush underwent a significant structural collapse as shown in Fig.~\ref{z}. Increasing the better solvent fraction by more than $x_{\text{b}}=0.1$ induced the swelling of the brush. The lowest and highest grafting densities achieve the minimum thickness at $x_{\text{b}}=0.05$, whereas the intermediate case reaches this at $x_{\text{b}}=0.1$. The main difference in the stimuli-response between the overlapping and the mushroom brush regimes is the reentrant transition to the swollen state. A first-order-like change appears in the overlapping brush, which show an abrupt change in the transition between collapsed and re-swollen states, whereas a continuous change is seen in the mushroom brush regime.

The brushes became more stretched in a pure better solvent than in a pure good solvent, consistent with the results of a previous study of a one-component solvent system.~\cite{Dimitrov07} In particular, a higher grafting density results in more drastic brush swelling. Accordingly, a larger difference in the thickness in the pure better solvent ($x_{\text b}=1.0$) was observed as the grafting density increased. Notably, the degree of collapse is most significant for intermediate grafting densities, where the thickness in the maximally collapsed state is reduced by half compared to the thickness in a pure good solvent. This suggests that there is an optimal grafting density at which the co-nonsolvency response, i.e., collapse transition, is maximized. We found that the range of solvent compositions where the brush thickness is lower than that in the pure good solvent becomes narrower as the grafting density increases. The co-nonsolvency region appears at $0.05\leq x_{\text b} \leq0.8$ for the lowest grafting density and $0.05\leq x_{\text b} \leq0.3$ for the highest grafting density. Here, we find that the window of polymer collapse strongly depends on the grafting density strongly as well as the temperature of the system,~\cite{Winnik90, Schild91} and the interaction asymmetry between the polymer and good solvent and polymer and better solvent.~\cite{Mukherji15, Mukherji16}

\begin{table}[t]
\small
  \caption{\ Comparison of the collapse ratio $h_{\text c}/h_{\text s}$ from AFM experiments (Exp.) and our simulations (Sim.). Here, the heights of the swollen brush $h_{\text s}$ and the collapsed brush $h_{\text c}$ correspond to the height at $x_{\text b}=0.0$ and $x_{\text b}=0.5$, respectively. The units of the experimental values of $\sigma_{\text g}$  and $h$ are [chains/$nm$] and [$nm$].}
  \label{AFM}
  \begin{tabular*}{0.48\textwidth}{@{\extracolsep{\fill}}llll}
    \hline
    $\sigma_{\text g}$ (Exp.) & $h_{\text c}/h_{\text s}$ (Exp.) & $\sigma_{\text g}$ (Sim.) & $h_{\text c}/h_{\text s}$ (Sim.)\\
    \hline
    0.03  & 0.25 & 0.002 & 0.55 \\
    0.27  & 0.32 & 0.074 & 0.64 \\
    0.69  & 0.50 & 0.148 & 0.90 \\
    \hline
  \end{tabular*}
\end{table}

To compare our simulation results with AFM experimental results~\cite{Sui11}, we calculated how the maximum brush height changes as the solvent is mixed with a co-nonsolvent, i.e., the collapse ratio. In AFM experiments, two overlapping brushes corresponding to $h \sim \sigma_{\text g}^{1/3}$, as well as one mushroom brush, composed of grafted PNIPAM chains were prepared. The brush height was measured in water/methanol mixtures of $50\text{ vol.}\%$, which corresponds to the reswelling composition of our simulations. The collapse ratio $h_{c}/h_{s}=h_{x_{\text b}=0.5}/h_{x_{\text b}=0.0}$, is shown in Table ~\ref{AFM}. Because our model does not target a specific co-nonsolvency system, but rather provides a general picture of diverse co-nonsolvency systems, we would not expect the collapse ratio to be identical absolutely. The difference between the values in the experiment and the simulation could arise from specific chemical details.~\cite{Scherzinger12} However, the decreasing trend of the collapse ratio with increasing grafting density is consistent with the experimental results. Interestingly, the relative change in the maximum brush height increases with increasing grafting density, but the relative change in the average height of monomers does not.

\newpage 

\subsection{Scaling behavior and the master curve in co-nonsolvency}

\begin{figure}[t]
\centering
  \includegraphics[width=8.5cm]{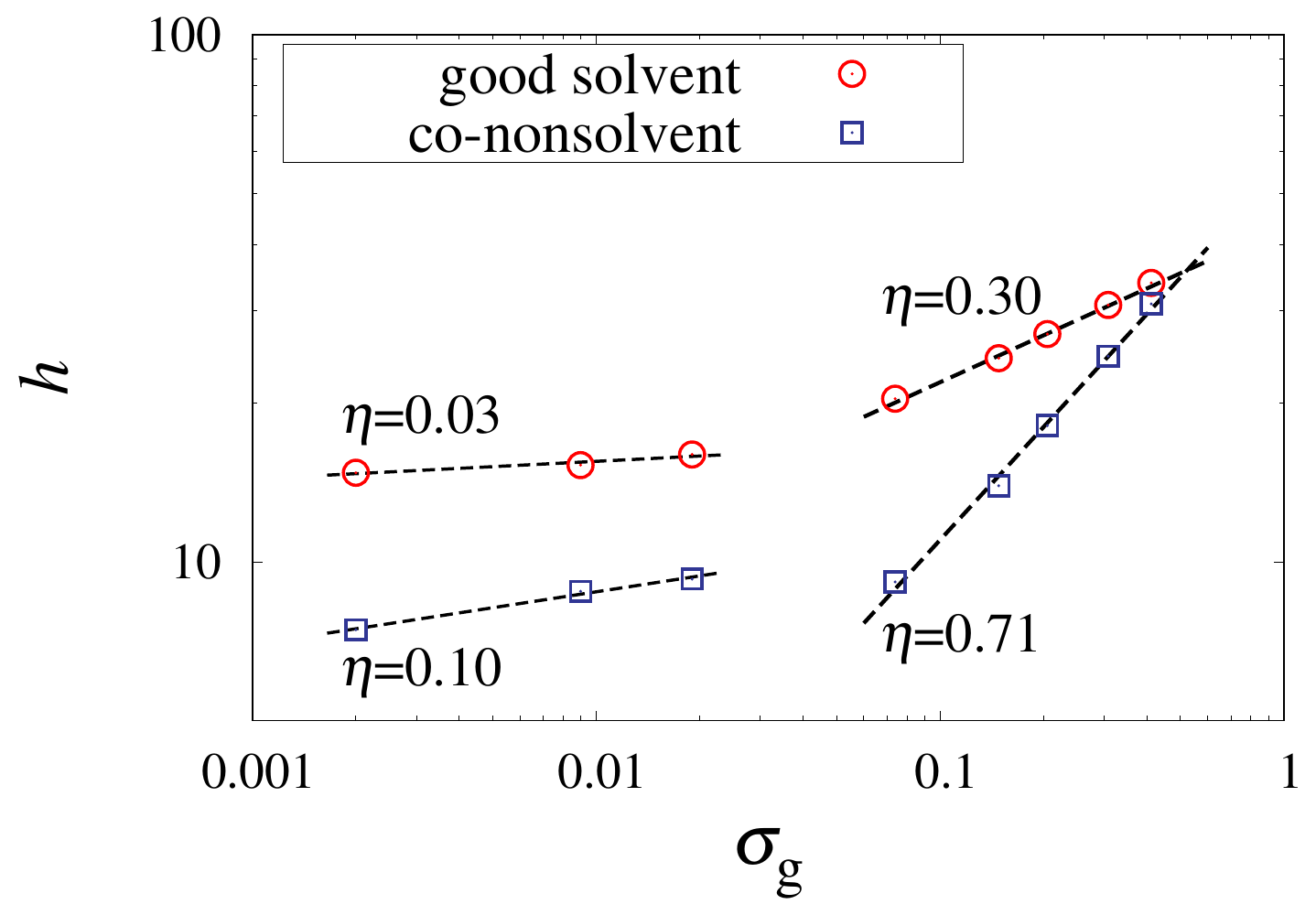}
  \caption{\color{blue}Log-log plot of maximum height as a function of grafting densities in the case of good solvent and co-nonsolvent environments. The dotted lines denote the power law $h \sim \sigma_{\text g}^{\eta}$ in the mushroom and the overlapping regimes. The scaling lines in the overlapping regime show drastic variation in their exponents in co-nonsolvent mixture ($\eta=0.30$ to $\eta=0.71$) whereas the mushroom regime does not ($\eta=0.03$ to $\eta=0.10)$. The value of $h$ for each grafting densities in good solvent mixture was chosen to obtain the maximally compacted co-nonsolvent fraction, as reported in the ESI${\dag}$.}
  \label{scale}
\end{figure}

For the brush height $h$, a power law for the brush in a good solvent $h \sim {\sigma_{\text g}}^{1/3}$ was developed by de Gennes and Alexander using a physically intuitive idea that considers a chain composed of subdivided blobs with a size equal to the distance between grafting sites.~\cite{Alexander77,deGennes80}. Recently, Moh \textit{et al.} reported, via experimental analysis, that overlapping brushes follow a scaling law $h \sim {\sigma_{\text g}} ^{0.8}$ in a poor solvent via experimental analysis.\cite{Moh11} For co-nonsolvents, however, the scaling behavior for the brush height has not been identified. Here, we found that the scaling exponent $\eta=0.71$ describes the height of the overlapping brushes as well, as shown in Fig ~\ref{scale}. As in the cases of good or poor solvents, the brush height in the co-nonsolvent follows two distinct scaling regimes in the mushroom and in the overlapping density range. The identified scaling exponents in a pure good solvent ($\eta=0.03$ and $0.30$ in the mushroom and brush regimes, respectively) are consistent with the results of previous experiments and theories.~\cite{deGennes80,Milner88} In the co-nonsolvent case, the scaling exponent in the brush regime changes more markedly from $\eta=0.10$ in the mushroom regime to $\eta=0.71$ in the overlapping regime, which is slightly less than that determined experimentally in the case of a poor solvent ($\eta=0.8$)~\cite{Moh11} and theory ($\eta=1.0$)~\cite{Halperin88}. The deviation of the scaling exponent from the value of standard poor solvent suggests that the brushes in the co-nonsolvent system are slightly more swollen than the collapsed brushes in the poor solvent where monomers are tightly packed. The exponent determined by us confirms the experimental observation; that is, a single PNIPAM chain globule still contains $\sim77\%$ solvent inside its hydrodynamic volume at a methanol volume fraction $x_{\text{methanol}}=0.2$.~\cite{Zhang01}
Another point to note is that as the grafting density increases, the maximum heights progressively become identical to each as the grafting density increases. 

Based on the scaling exponent ($\eta=0.71$) reported here and the Alexander-de Gennes brush theory,~\cite{Alexander77,deGennes80} we can construct a monomer density profile analytically. Because collapsed brushes in a poor solvent have shown rectangle shaped profiles in the previous studies,~\cite{Grest44, Dimitrov07} we can assume the monomer density curve in the co-nonsolvency environment has a the rectangular shape,
\begin{equation}
\rho_{\text m}(z) \sim \Theta(h-z). 
\label{eq:master1}
\end{equation}

\begin{figure}[t]
\centering
  \includegraphics[width=8.7cm]{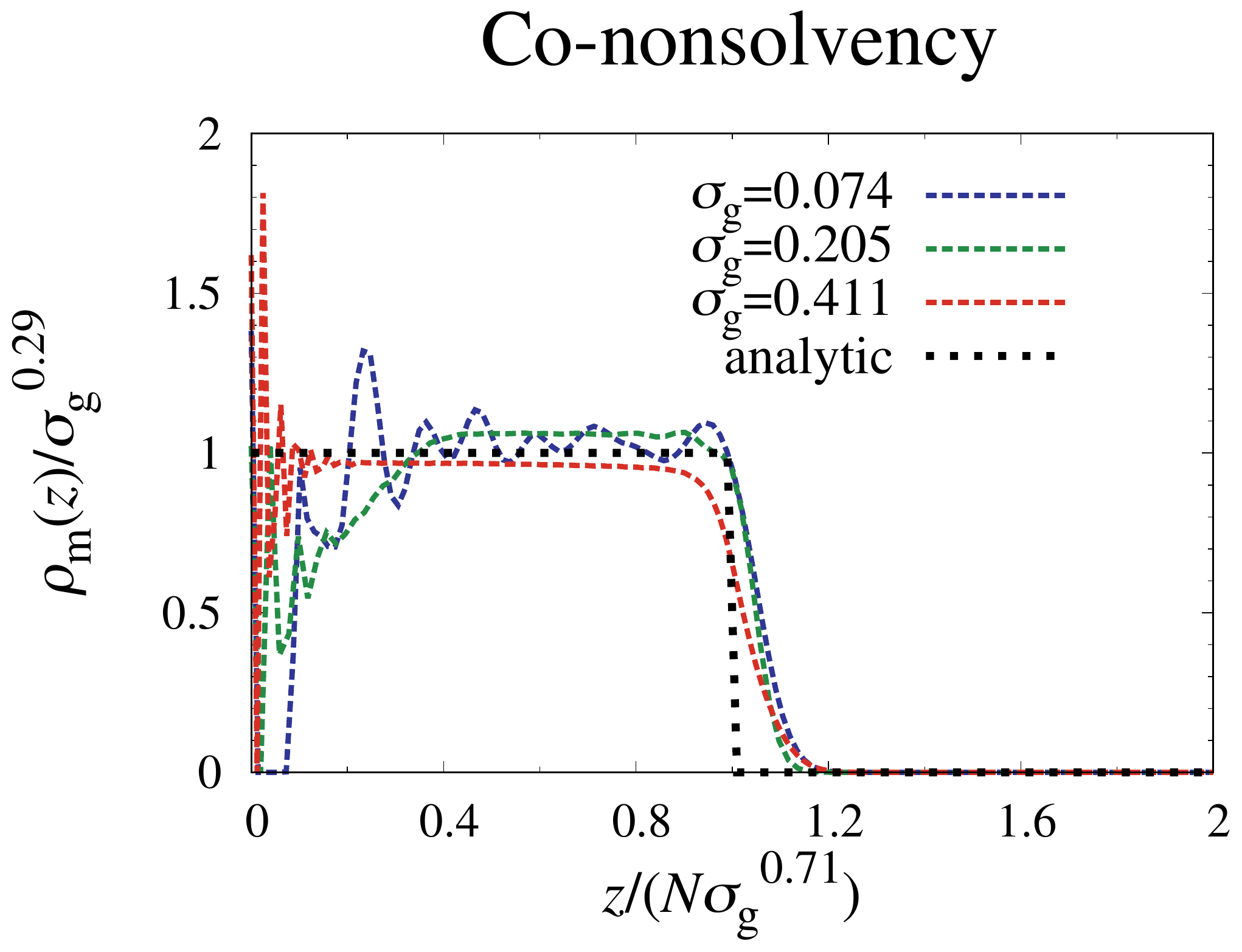}
  \caption{\color{blue}Analytic master curve and the numerical density profiles of monomers. The black analytic function defined as Eq. \ref{eq:master5} is based on an exponent value $\eta=0.71$ from the numerical data. The colored density curves represent simulation data of three overlapping brushes with grafting densities $\sigma_{\text g} =0.074, 0.205, 0.411$. The density profiles for each brush are the most compact profiles in the co-nonsolvency regime, which correspond to $x_{\text b}=0.15, x_{\text b}=0.05$, and $x_{\text b}=0.1$ at $\sigma_{\text g} =0.074, 0.205$, and $0.411$, respectively.}
  \label{master}
\end{figure} 

For the grafting density $\sigma_{g}$, the average distance between the grafting sites $D$ can be expressed as $D=\sigma_{\text g}^{-1/2}$. Using Alexander-de Gennes brush theory, the average distance is equal to the size of correlation blobs, and the number of correlated monomers in the correlation blob, $g_{D}$ is related to the grafting distance by $D =b{g_{D}}^\nu$, where $b$ denotes the monomer length and $\nu$ denotes the Flory exponent in the co-nonsolvent. On the scale of the correlated volume $D^3$, the monomer density in the brush region ($z<h$) can be written as
\begin{equation}
\rho_{\text m}(z) \simeq \frac{g_D}{D^3}=b^{-\frac{1}{\nu}}D^{\frac{1}{\nu}-3}=b^{-\frac{1}{\nu}}\sigma_{\text g}^{\frac{1}{2}(3-\frac{1}{\nu})}.
\label{eq:master2}
\end{equation}
In terms of the whole chain, the monomer density can be expressed in terms of the number of monomers $N$, and the brush height $h$ as
\begin{equation}
\rho_{\text m}(z) \simeq \frac{N}{hD^2}=\frac{N}{h{\sigma_{\text g}}^{-1}}.
\label{eq:master3}
\end{equation}
Equating Eq. ~\ref{eq:master2} into Eq. \ref{eq:master3} results in 
\begin{equation}
h=N b^{-\frac{1}{\nu}} \sigma_{\text g}^{\frac{1}{2\nu}-\frac{1}{2}}.
\label{eq:master4}
\end{equation}
Here, using $h \sim {\sigma_{\text g}}^{0.71}$ from the simulation, we can obtian the Flory exponent $\nu \simeq 0.413$. Finally, by substituting this Flory exponent $\nu \simeq 0.413$ into Eq. \ref{eq:master2} and Eq. \ref{eq:master4}, we obtain the monomer density function in the form of
\begin{equation}
\rho_{\text m}(z)=b^{-2.42} \sigma_{\text g}^{0.29}\Theta(N b^{2.42} \sigma_{\text g}^{0.71}-z).
\label{eq:master5}
\end{equation}
In our system, the monomer length is given by $b=1.0\sigma$. Fig.~\ref{master} shows that the above scaling argument works well in that the density profiles fall onto a single master curve with appropriate rescaling. Note that there is no fitting parmater in Fig.~\ref{master}. By the combination of simple theory and the numerical plots, we successfully obtained a the novel function that predicts the monomer density in the co-nonsolvent environment.

\subsection{Density profiles}

\begin{figure}[t!]
\centering
\includegraphics[width=13.0cm]{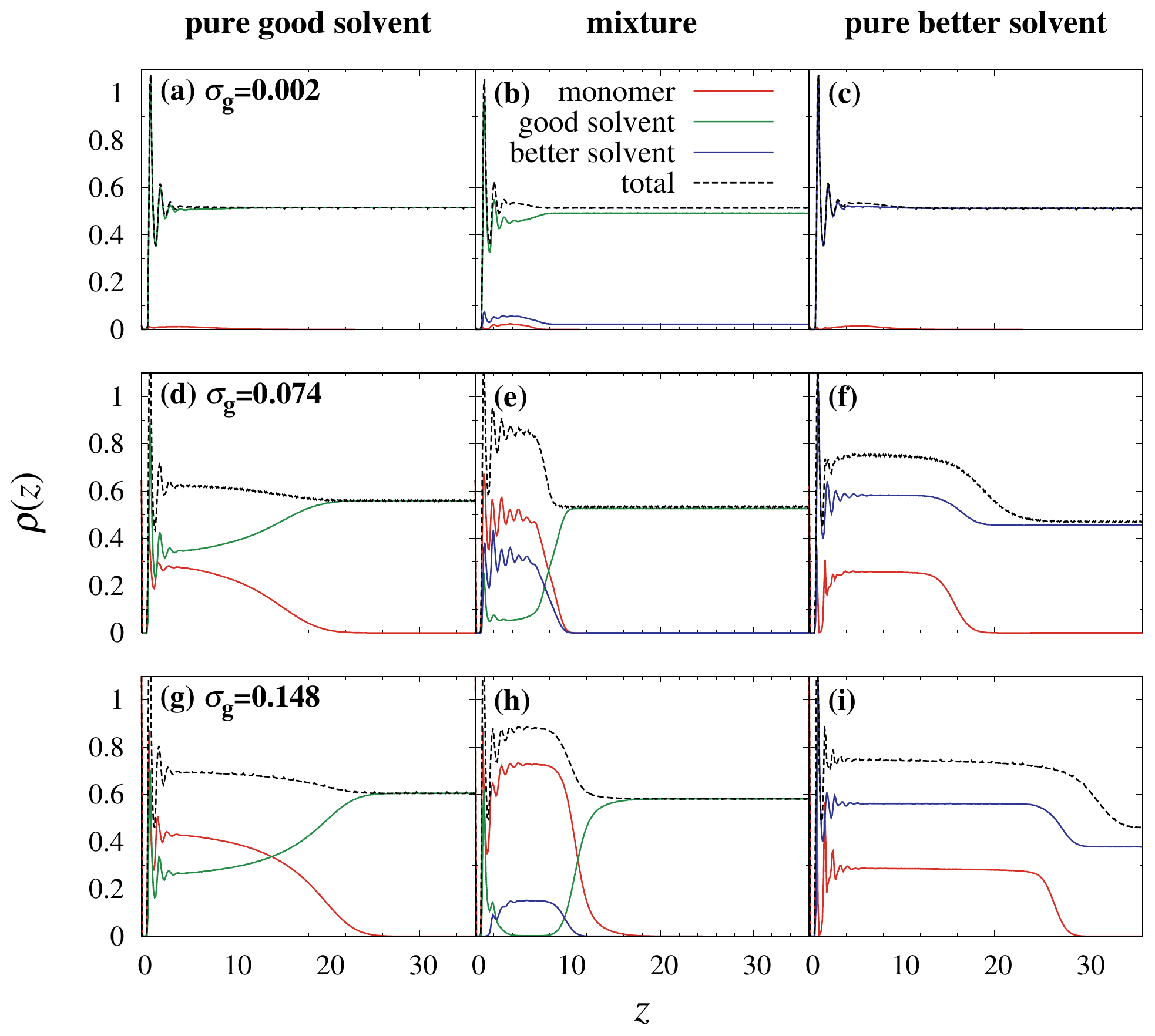}
\caption{Density profiles of the monomer, good solvent, better solvent, and total particles (red, green, blue, and black dotted lines, respectively) as a function of the vertical coordinate z in the pure good solvent (left), mixture (middle), and the pure better solvent (right). The total density was calculated as the sum of the monomer, good solvent, and better solvent densities. Graphs are shown for three different grafting densities, i.e., the lowest ((a),(b),(c)), the intermediate ((d),(e),(f)), and the highest grafting density ((g),(h),(i)).The insets show magnified density profiles for the case of the lowest grafting density.} 
\label{tot_density}
\end{figure}

\begin{figure}
\centering
\includegraphics[height=4.2cm]{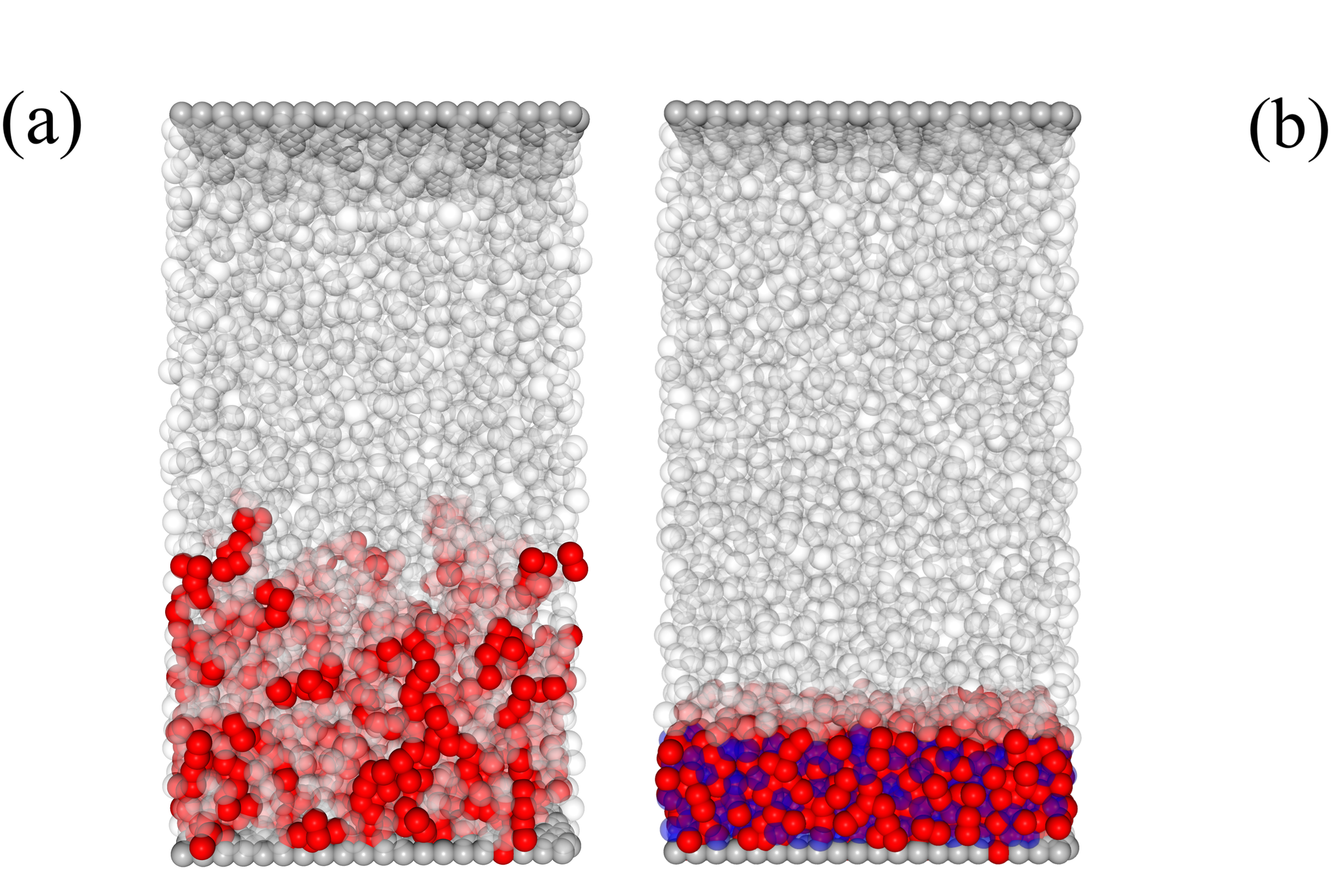}
\includegraphics[height=3.2cm]{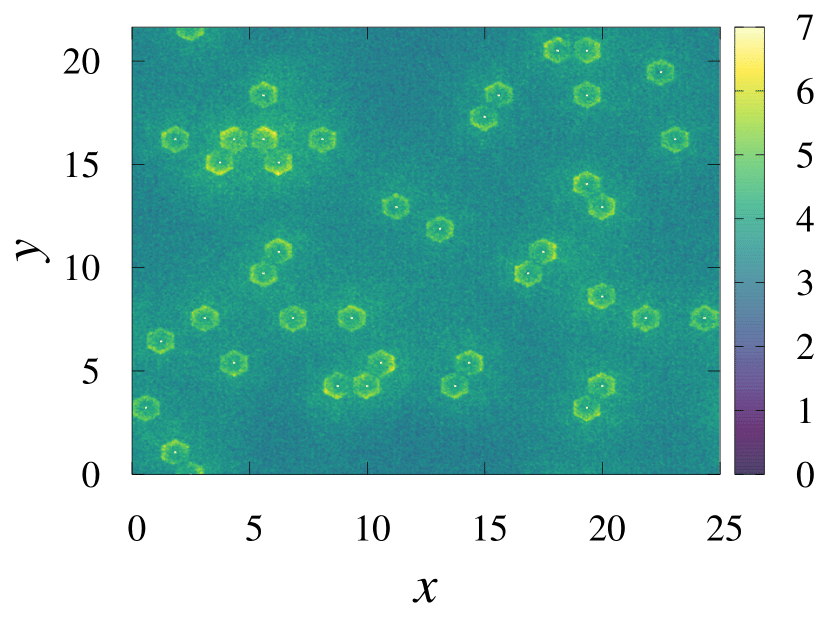}
\includegraphics[height=3.2cm]{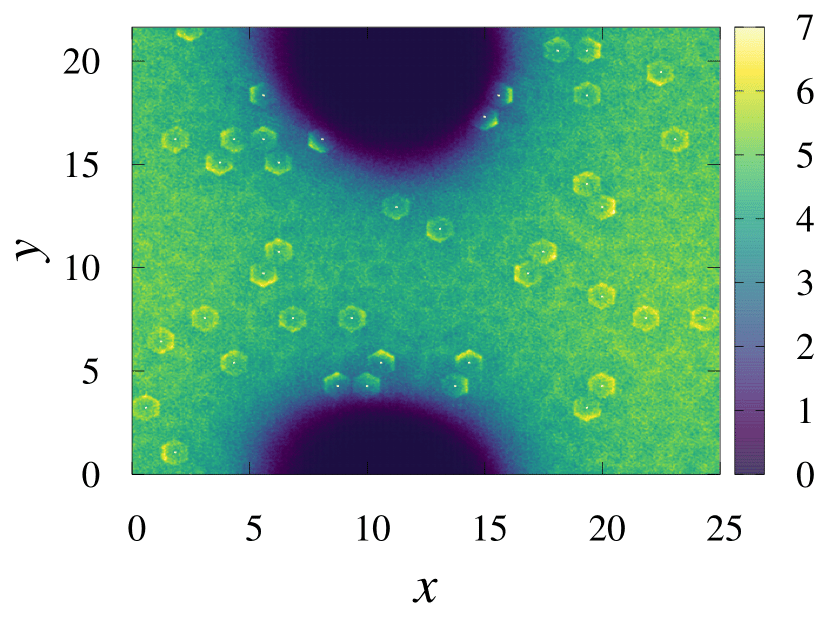}
\label{P40_0_0f}
\caption{\color{blue}(a) Simulation snapshot in a pure solvent (left) and in co-nonsolvents mixture ($x_{\text b}=0.1$) (right), and (b) average number of monomers per surface area in the pure good solvent (left) and the mixture (right) at the grafting condition $\sigma=0.074$. In (a), the transparent, red and purple spheres represent good solvent, monomer and better solvent, respectively. Good solvent phase is separated from the polymer/better solvent phase at $x_{\text b}=0.1$. In (b), The yellow points represent the grafted monomers in each chain. Hexagonal structures in the both of plots are formed due to the restrained second monomers which are bonded to the fixed first monomers.}
\label{brush_figure}
\end{figure}

Motivated by the observation of the transition in brush thickness at various solvent compositions, we determined the overall structure of the system (Fig.~\ref{tot_density}). The density profiles of the monomers in the pure solvent and the mixture where each brush is maximally collapsed were calculated. The distribution of the solvent molecules causing the structural change of the brush is also shown. For the lowest, intermediate, and highest grafting density, the brush collapse is most significant at $x_{\text b}=0.05, 0.15$, and $0.1$, respectively. Those for the rest of the compositions are shown in Fig. S2 in the ESI${\dag}$.

For the intermediate and highest grafting densities, a parabolic density profile with an exponential tail was observed in the pure good solvent environment, as shown in Fig.~\ref{tot_density}(d). In the mixture, the monomers are highly concentrated near the grafting surface, forming a narrow brush regime. In the case of the solvent mixture, the density profile exhibits a steep descent at the interface, which is consistent with the density profiles in poor solvents observed in previous studies.~\cite{Grest44, Dimitrov07} In this solvent regime, a wide range with a uniform density was obtained, and the density profile decreases sharply at the interface. Note that the parabolic curve re-emerges gradually as the better solvent fraction increases. For the lowest grafting density, in contrast, the density profile maintains a Gaussian-like shape over the whole range of solvent compositions.

With the addition of the better solvent molecules, layered structures are observed prominently in the overlapping brushes, which are typically observed in confined systems such as ionic liquids in the electric double layers.~\cite{Jo17}  The first layers with sharp peaks near the grafting surface were observed regardless of the solvent composition. Such a highly ordered monomer structure is due to the geometric constraints induced by the repulsive interaction with the grafting wall. The ordering caused by topological restrictions persists for up to about three layers in pure solvent conditions for all three grafting densities. However, in the solvent mixture, layered structures extend out over a significantly longer distance in the vertical direction. In particular, for the brushes with an intermediate grafting density, the ordering remains until the monomers reach the interface between the brush and bulk in the mixture. This indicates that the monomers are densely packed throughout the whole range of brush regions for the intermediate grafting density, which is supported by the disappearance of an exponential tail structure near the interface and the higher total density $\rho_{\text t}$ in the brush region than bulk.

In the pure good solvent conditions, the good solvent exists in the brush region and in the bulk. In the mixture, however, the good solvent molecules near the grafting polymers are pushed towards the bulk region. In particular, the good solvent molecules are scarce near the grafted polymers at intermediate and highest grafting densities. The grafting brush with intermediate grafting density shows that both the good solvent and the better solvent exist in the brush region in the mixture because the better solvent molecules form clusters in regions where the monomer contact probability is high. As shown in Fig.~\ref{brush_figure}, in the polymer-free zone where the local grafting is relatively low, the better solvent does not gather to form chain-better solvent-chain aggregates and the good solvent is in contact with the surface.

The density profiles for the better solvent show the opposite trend to that of the good solvent. In the mixture, the better solvent is locally concentrated near the monomer-rich region. This is clearly different behavior from that observed in the poor solvent where the monomers aggregate with each other alone and solvent molecules are excluded. This may explain the different scaling behavior in the co-solvent mixture and the poor solvent (Fig.~\ref{scale}). For the lowest grafting density, the change in the density in the better solvent is smaller than that of the overlapping brushes. For this grafting condition, the better solvent remains near the grafting surface at all solvent compositions. However, the accessibility of solvent molecules to monomers increases as the polymer conformation changes during swelling. In this context, a recent QCM experiment is worth mentioning. The results of the QCM experiments suggests that the solvent-polymer contact area in the methanol/water compositions indirectly from the mass change in the polymer growth process.\cite{Pomorska16} At a methanol fraction $x_{\text m}=0.1$, the heterogeneous growth occurs in grafted chains, leading to a non-uniform length distribution, whereas, at $x_{\text m}=0.16$, the homogeneous growth was observed. Combining our density profile and the QCM observations, we can argue that monomers have access to the active radical sites on the polymers only at the interface between the brush and the brush at $x_{\text m}=0.1$ (Fig.~\ref{tot_density}(h)), whereas all the chains are solvated with the better solvent at a methanol composition greater than $x_{\text m}=0.1$ (see Fig. S2 in the ESI${\dag}$). Thus, our density profile gives direct information concerning "grafted from" polymerization, which could be important in controlling the dispersity of polymer size in the polymerization process.   

\begin{figure}[ht!]
\centering
 \includegraphics[width=8.0cm]{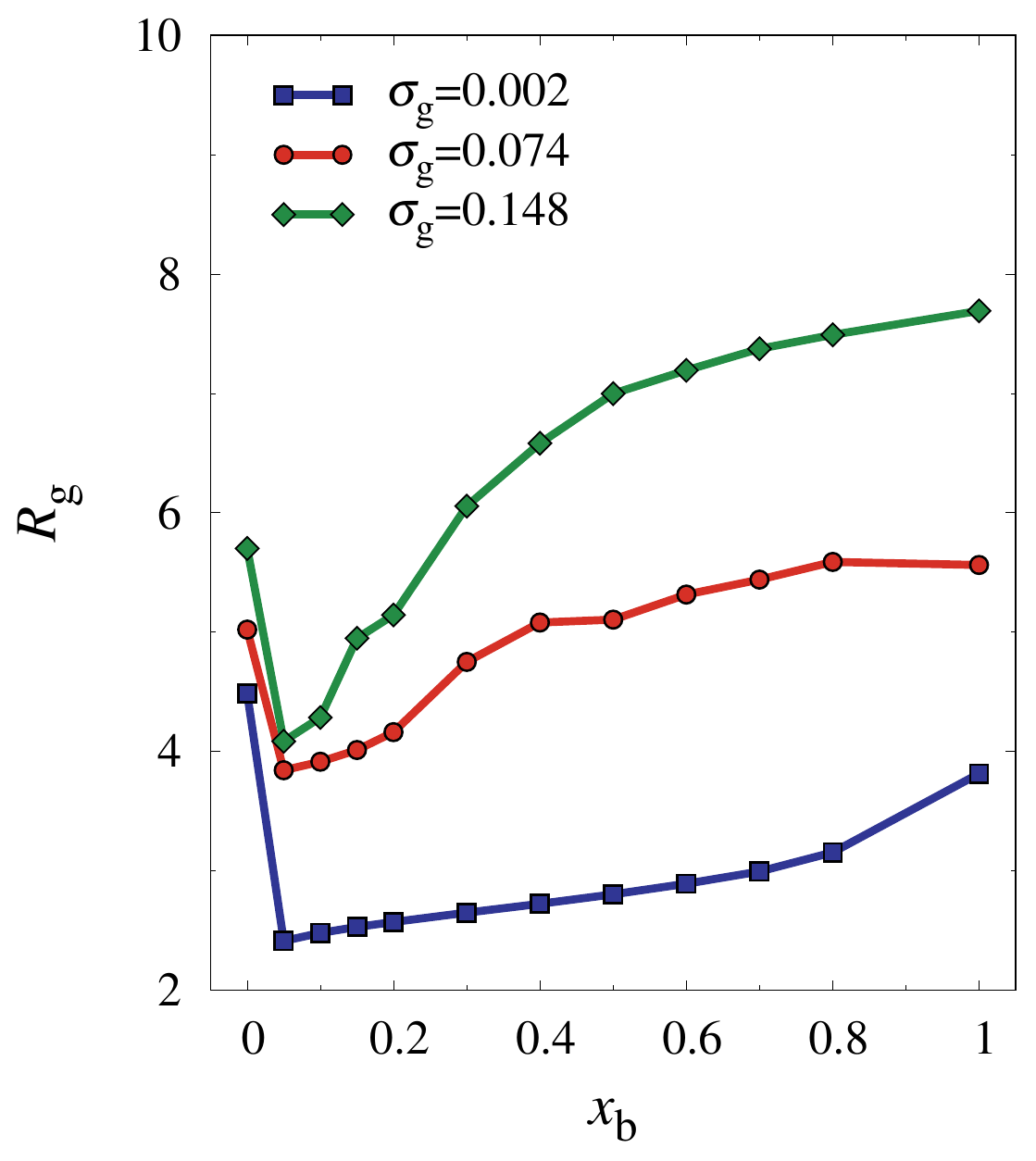}
\caption{Radius of gyration as a function of the better solvent fraction for three grafting densities.}
\label{Rg}
\end{figure}

The change in the total density shows a different behavior in the brush region and in the bulk. In the mixture, the concentration of the particles increases in the brush region and slightly decreases in the bulk compared to the good solvent condition. The total density of brush region exceeded $\rho\simeq 0.8$, at which the better solvent and the monomer behave like a compact packed crystal. The observed heterogeneity in the density and the local concentration of the better solvent support the previous result that co-nonsolvency is the energetically dominant phenomenon.~\cite{Mukherji14}

\subsection{Conformation and orientation of polymers}

To investigate the structural properties of the polymer brushes more thoroughly, we extended our analysis further for the structural properties of individual polymers, such as the radius of gyration and the orientation of polymers. Ensemble-averaged values of each property were calculated by sampling over all polymer configurations and time-steps. First, we calculated the radius of gyration by varying the solvent composition for the three grafting densities (Fig.~\ref{Rg}). For all cases of grafting densities, the radius of gyration takes the minimum value at $x_{\text{b}}=0.05$ as does the brush thickness of the brush (Fig.~\ref{z}). However, the change in the size of the individual polymer is not consistent with the change in the thickness. Even though the polymer clusters collapses to the greatest extent in the intermediate grafting density system, the polymer chain itself shrink the most in the lowest grafting density system.

\begin{figure}[t!]
\centering
 \includegraphics[width=8.0cm]{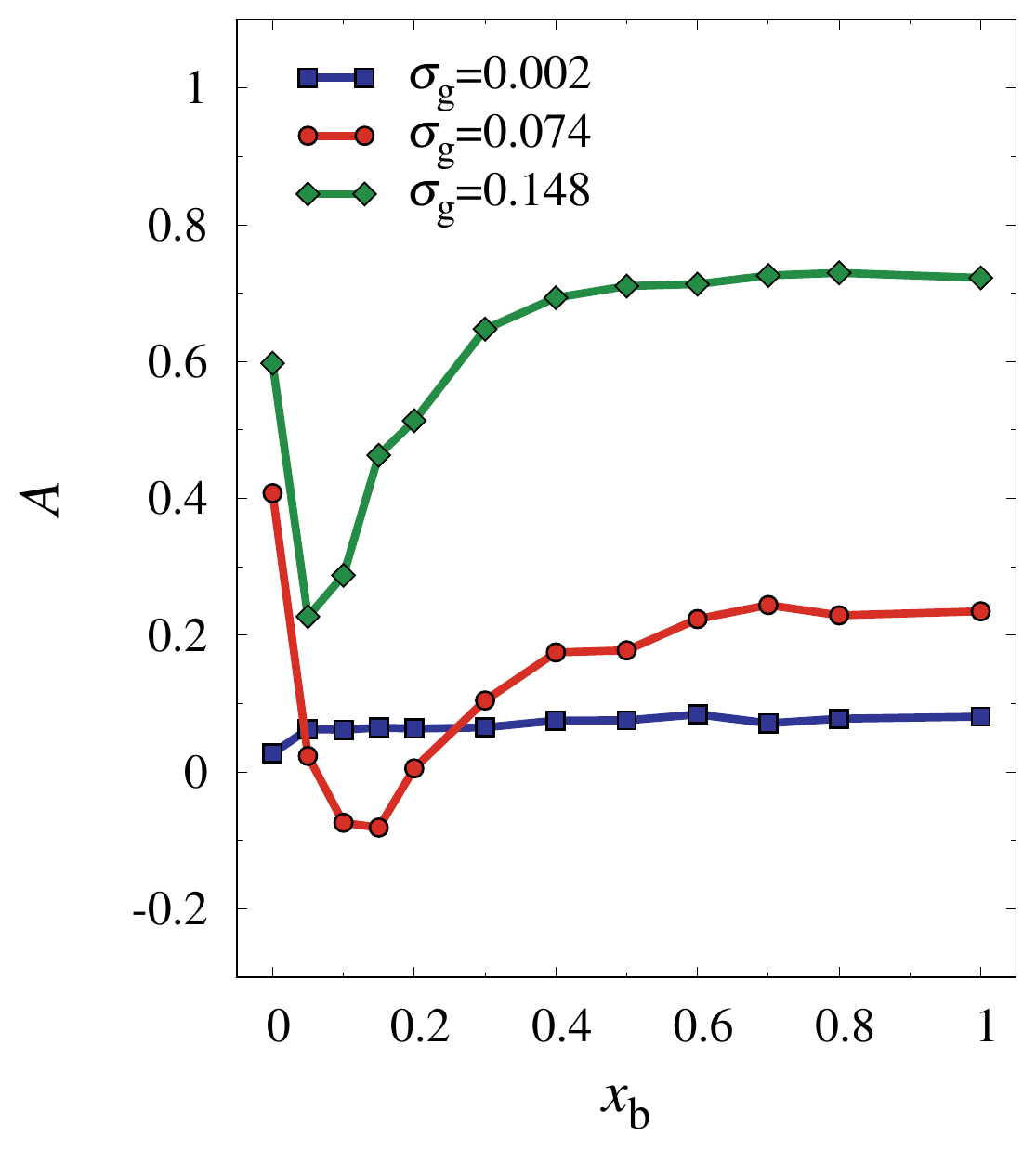}
\caption{Orientational parameter $A$ as a function of the better solvent fraction. The orientational parameter defined in Eq.~(\ref{eq:A}) is averaged over the polymer chains and the time.}
\label{Rg_conf}
\end{figure}

A more detailed view of the structural change of the polymers can be obtained by investigating the orientational properties. Thus, we introduce an orientational order parameter $A$, which is defined as the ensemble average of the anisotropy in the 
gyration radius vector,~\cite{Pastorino06}  
\begin{equation}
A=\bigg\langle \frac{{R_{{\text g},z}}^2-(1/2)({R_{{\text g},x}}^2+{R_{{\text g},y}}^2)}{{R_{{\text g},x}}^2+{R_{{\text g},y}}^2+{R_{{\text g},z}}^2} \bigg \rangle, \label{eq:A}
\end{equation}
where $R_{{\text g},x}$, $R_{{\text g},y}$, and $R_{{\text g},z}$ denote the Cartesian components of the radius of gyration vector, defined by
\begin{equation}
R_{{\text g},\alpha}^2 = \frac{1}{N}\sum_{i=1}^{N}({\bf R}_{i,\alpha}-{\bf R}_{\text{CM},\alpha})^2
\end{equation}
with the center of mass vector ${\bf R}_\text{CM}$ and $\alpha=x, y,$ and $z$. 
From this definition, one can note a positive value of $A$ characterizes a perpendicular orientation to the grafting surface, a negative value indicates that the polymer is oriented parallel to the substrate, and a negligible value of $A$ indicates an isotropic polymer conformation. For the intermediate grafting density, the value decreases from positive to negative, indicating that the orientation of polymers changes from the perpendicular to the parallel direction (Fig.~\ref{Rg_conf}). The polymer prefers the in-plane orientation the most at $x_{\text{b}}=0.15$, but is reoriented to the perpendicular direction at $x_{\text{b}}=0.4$, suggesting that the collapsed structure is caused by entanglement of the polymers aligned in the $xy$ plane. In contrast, for the lowest and highest grafting densities, the values are positive and zero, respectively, which means that polymers maintain a similar orientation over the whole range of better solvent fractions.

\begin{figure}[t]
\centering
 \includegraphics[width=8.0cm]{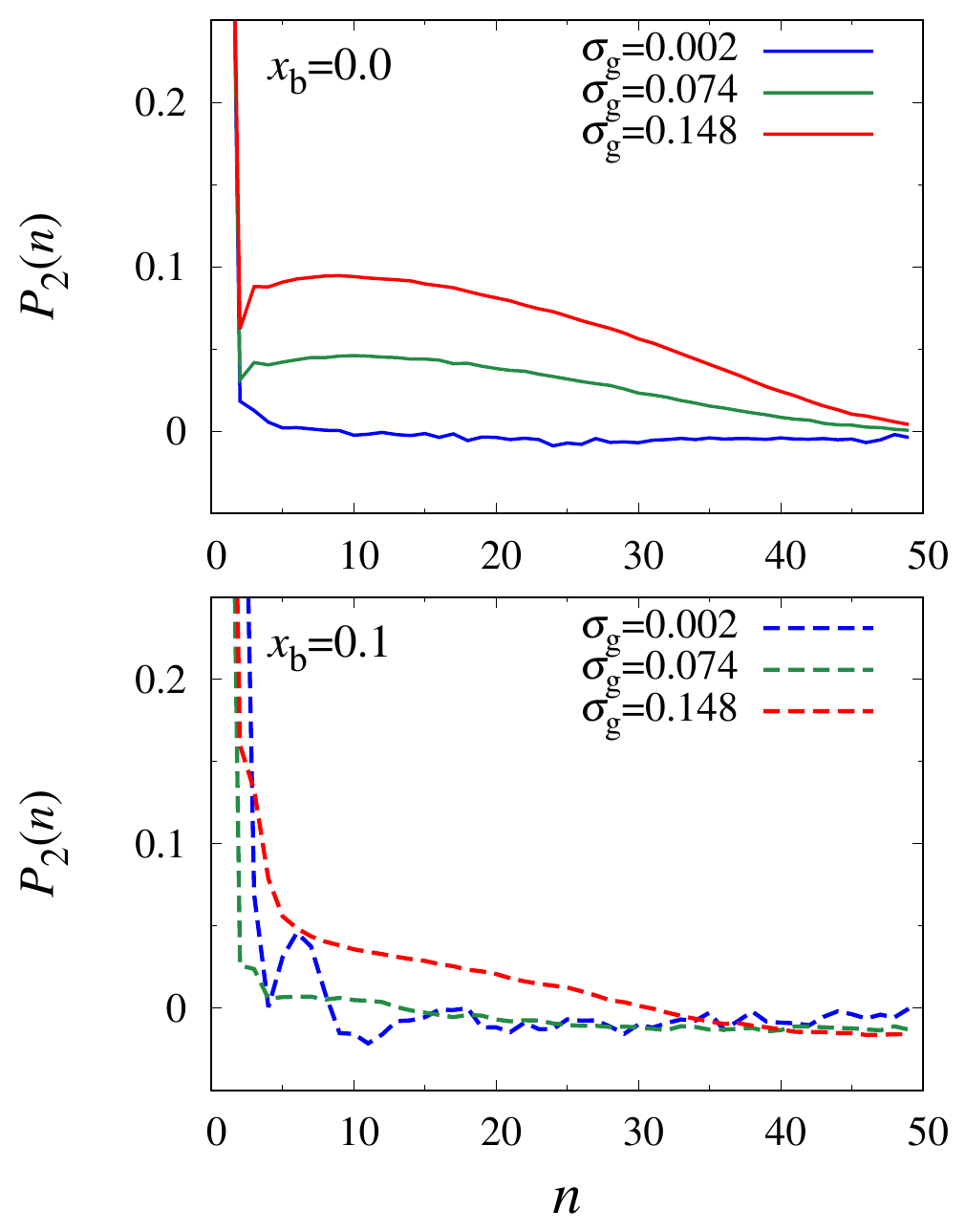}
\caption{Second Legendre polynomial as a function of the bond sequential number for the three grafting densities. The total number of bonds in a chain is $n_{\text{bond}}=49$. The solid line represents the pure good solvent environment, and the dotted line represents the distribution in the solvent environment when $x_ {b} = 0.1$. Blue, green, and red lines represent the lowest, intermediate and the highest grafting density conditions, respectively.}
\label{bond_ori}
\end{figure}

We obtained the bond orientation of polymer as a function of the monomer step $n$ via the second Legendre polynomial 
\begin{equation}
P_{2}(n)=\frac{3}{2}\langle \cos^{2} \theta_{n} \rangle-\frac{1}{2} \label{eq:Pn},
\end{equation}
where $\theta_{n}$ denotes the angle between the bond vector with the bond sequential number $n$ and the normal vector to the grafting surface.(Fig.~\ref{bond_ori}) A random alignment of the bond without any orientational preference converges the value of $P_{2}(n)$ to zero. A positive value of $P_{2}(n)$ indicates a preference for the perpendicular orientation whereas a negative $P_{2}(n)$ indicates the parallel orientation. For an initial step of $n=1$, the observed values of $P_{2}(n)$ are greater than 0.6 because of all conditions due to the grafting effect. In a pure good solvent, this preference for the vertical direction disappears quickly in the initial steps in the case of the low grafting density, and a self-avoiding walk is observed. On the other hand, in the case of the intermediate and high grafting density, the perpendicular orientation of the brush was maintained by the emergence of the excluded volume effect between the chains. This originates from the fact that the walk of the bonds is disturbed in the parallel direction by repulsive interactions with other chains. Because better solvent molecules make multiple contacts with the monomers in the mixture, the interaction between the monomers becomes attractive. When the co-nonsolvent is introduced to the mushroom brush system, the chains forms a globule structure to encase the better solvents in monomers, which is represented by the damped oscillation of the curve.

\subsection{Segmental properties: Inter or Intramolecular bridging}

\begin{figure*}[t!]
\centering
 \includegraphics[width=16.5cm]{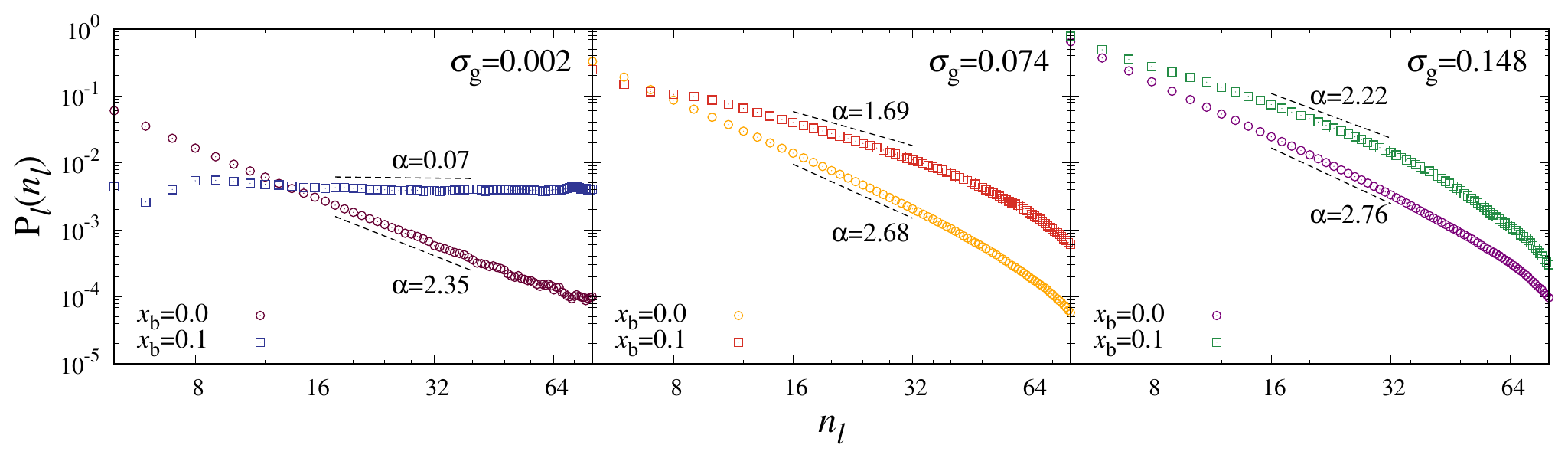}
\caption{\color{blue}Log-log plot of probability distributions of loop size $n_l$ at the lowest(left), the intermediate (middle), and the highest grafting density brush(right) with solvent fractions $x_ {\text b} = 0.0$ and $x_ {\text b} = 0.1$. The power-law exponent $\alpha$ was obtained by fitting the simulation data.}
\label{loop}
\centering
 \includegraphics[width=16.0cm]{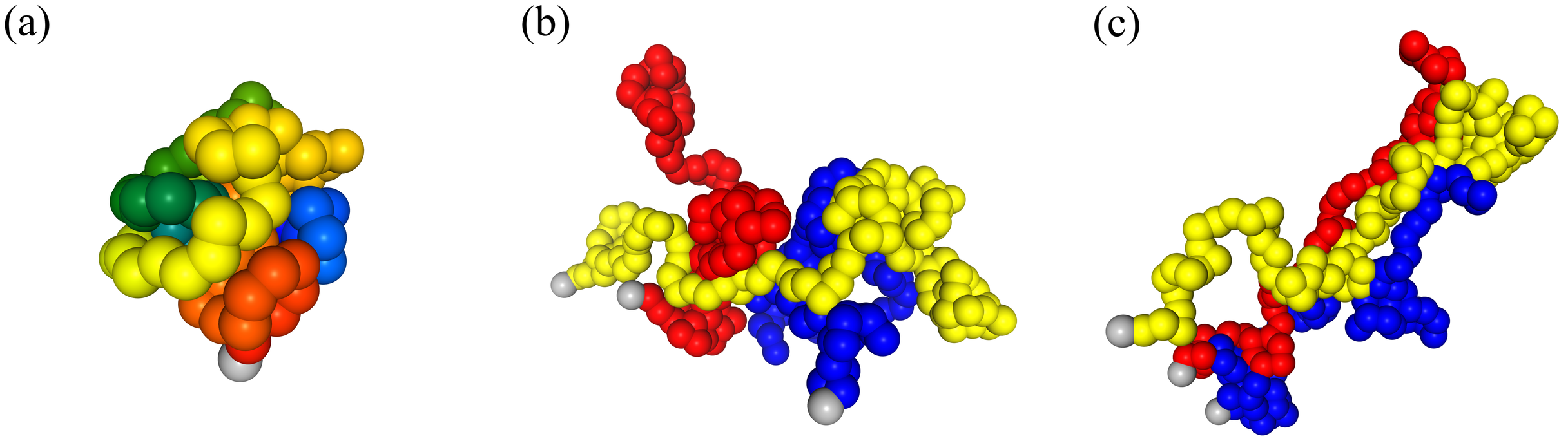}
\caption{\color{blue}Simulation snapshot of looping configuration for (a) the lowest, (b) intermediate, and (c) the highest grafting density brush. The color for the lowest grafting density brush varies with the monomer index in rainbow sequence. For the intermediate and the highest grafting density brushes, three different chains are represented by the different colors. The two configurations in (a) correspond to the side view and bottom view of same configuration. Only the polymer of the mushroom brush formed a globular configuration because of the intramolecular bridging, whereas the overlapping brush polymers show more stretched configurations when the looping and unlooping structures coexist.}
\label{loop_fig}
\end{figure*}

Increasing the excluded volume between chains by increasing the grafting density changed various structural properties such as the thickness, orientation, and the sequential bond directions in the co-nonsolvent. We attribute these diverse structural properties to the different preferences for interbridging and intrabridging structures of the polymers with different grafting densities. Mukherji \textit{et al.}, who first calculated intrabridging structures in a single polymer system, considered a bridging structure to have formed if a better solvent molecule makes contacts with two or more monomers in a chain simultaneously. However, for the crowded polymer systems such as polymer brushes, the calculation of bridging events may not be straightforward. Instead, we calculated the looping and unlooping events to distinguish intramolecular intermolecular bridging structure for each grafting density. The probability distribution of monomer loop with length $n_\textit{l}$ is defined as
\begin{equation}
P_{\textit l}(n_{\textit l}) = \bigg \langle \frac{1}{N-n_{\textit l}} \sum_{i=1}^{N-n_{\textit l}} \Theta \big(r_{\text c}-r(i,i+n_{\textit l})\big) \bigg \rangle 
\label{eq:loop},
\end{equation}
where $N$ denotes the number of monomers in a chain and $i$ represents the monomer index. $r(i,i+n_{\textit l})$ is the distance between two monomers separated by length $n_{l}$. When $r(i,i+n_{\textit l})$ is less than the cut-off distance $r_{\text c}$, taken as 1.5$\sigma$ in this work, a loop of length $n_{l}$ is considered to be formed. Similarly, the unlooping probability is defined as 
\begin{equation}
P_{\textit ul}(n_{\textit ul}) = \bigg \langle \frac{1}{N-n_{\textit ul}} \sum_{i=1}^{N-n_{\textit ul}} \Theta \big( r(i,i+n_{\textit ul}\big) - r_{\text c})  \bigg \rangle.
\label{eq:unloop}
\end{equation}
where $\langle \dotsb \rangle$ means that probability is averaged over for the chains and the time. The probability of looping and unlooping was calculated for the segments with at least size $n_{\textit l}, n_{\textit{ul}} \geq 5$. From Eq. \ref{eq:loop} and Eq. \ref{eq:unloop}, the normalization is satisfied as $P(n_{\textit l})+P(n_{\textit{ul}})=1$ when $n_{\textit l}=n_{\textit ul}$. Generally, the looping probability follows a power-law scaling behavior for a loop of size $n_l$ ($P(n_l)\sim n_l^{-\alpha}$) where the exponent value $\alpha$ varies by external or internal factors such as solvent environment or chain stiffness. As shown in the Fig.~\ref{loop} we see the simple power law in the mushroom case ($\sigma_{\text g}=0.002$). The looping probability of mushroom brush in the co-nonsolvent mixture indicates that a nearly uniform distribution was observed having significantly small exponents ($\alpha \simeq 0.07$) compared to the standard poor solvent one ($\alpha \simeq 1.0$ or $1.6-1.7$).\cite{Toan08,Debnath04} This implies that the folding in each solvent induces remarkably different looping behavior, although both the poor solvents and co-nonsolvent mixture induce a globular polymer structure. Single polymer folding in the co-nonsolvent takes place as follows: monomers capture and surround the better solvent, whereas the monomers in the normal poor solvent simply maximize the contact between monomers. Although the mushroom brush system has a power law exponent independent of the loop size, variations in the tangential line slopes were obtained for the overlapping brush systems ($\sigma_{\text g}=0.074$ and $0.148$). Actually, this behavior is consistent with the Alexander-de Gennes brush theory, where the subunits $j=1,\dotsb,N/g_{\text D}$ are stretched in the vertical direction from the surface. Based on the Alexander-de Gennes brush theory, the monomers in the subunit $j$ do not meet the monomers in the subunits $1,\dotsb,j-1$, prohibiting the looping between monomers in different subunits. Because the configurations of monomers coming back to the previous subunits are excluded, the looping probability decreases as the loop size increases, meaning that the looping exponent is dependent on the loop size nl. This argument is supported by the fact that a constant tangent value is sustained up to $n_{l}=g_{\text D} \simeq b^{-1/\nu} D^{1/\nu} \simeq b^{-1/\nu} {\sigma_{\text g}}^{-1/2\nu}$ (Fig. S10 in ESI${\dag}$), where only the self-avoiding walk affects to the loop conformation. Therefore, we obtained the characteristic exponent $\alpha$ in the overlapping brush for a representative region, $16 \leq n_l \leq 32$. The scaling exponent in the case of mushroom brush in a good solvent is close with values for the free polymer in a good solvent which was obtained in the previous studies($\alpha \simeq 2.2-2.4$).\cite{Podtelezhnikov97,Debnath04,Thirumalai99,Toan08} The exponents in overlapping brush($\alpha \simeq 2.68$ and $2.76$) are larger than those of the mushroom brush because the preference for the stretched conformation of the chain increases as the neighboring chains increase. The gap between exponent values in pure good solvent and co-nonsolvent mixture decreases as the grafting density increases. As shown in Fig.~\ref{loop_fig}, the abundance of neighboring chains induces the intermolecular bridging rather than the intramolecular bridging structures. The increasing frequency of intermolecular bridgingevents increases the inhibition of self-interacting loop conformation. Consequently, small loops are more dominant at the intermediate grafting density than the high grafting density. The unlooping probabilities $P(n_l)$ is provided in the ESI${\dag}$.

\begin{figure*}[t!]
\centering
 \includegraphics[width=8.0cm]{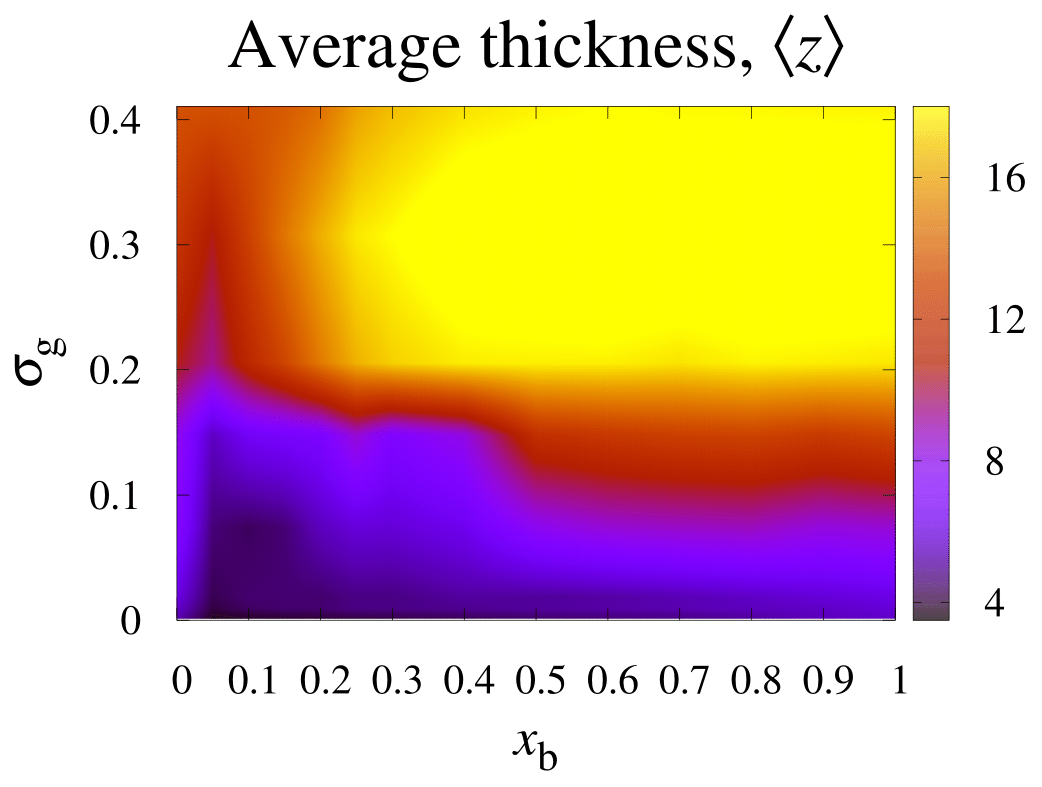}
 \includegraphics[width=8.0cm]{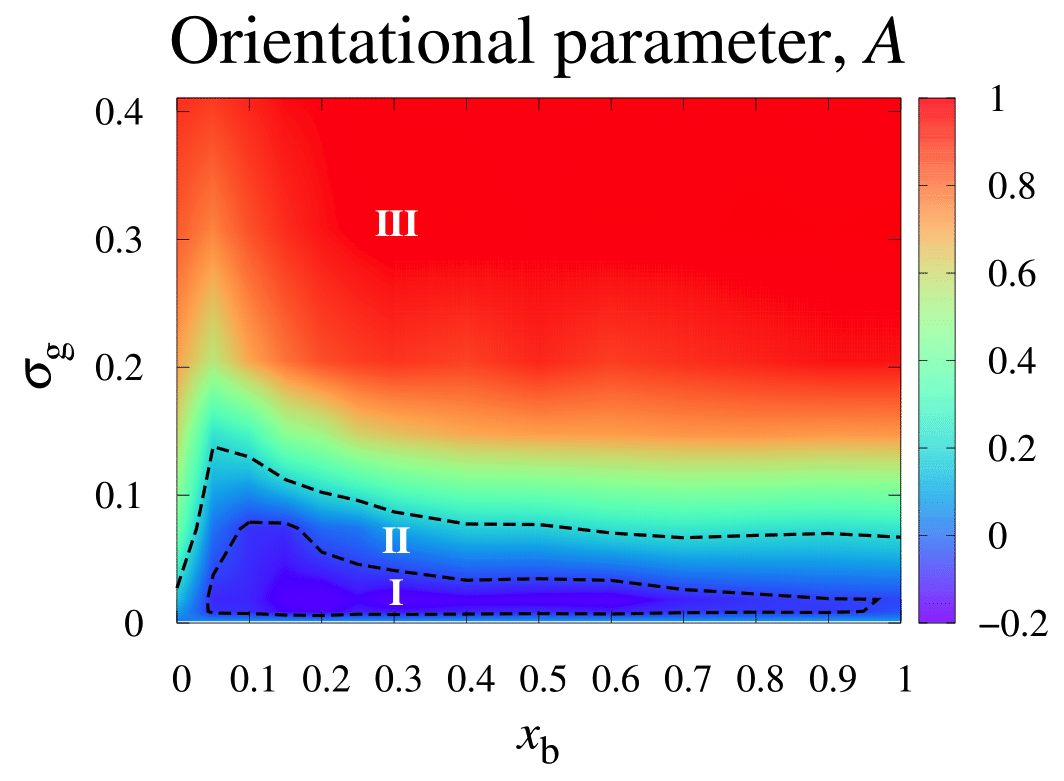}
\caption{Phase diagram for the average thickness  $\left<z\right>$ (above) and the orientational order parameter $A$ (below) as a function of the better solvent fraction and the grafting density. The two dotted lines correspond to $A = 0.0$ and $0.2$ and divide the parallel(I), isotropic(II) and perpendicular(III) orientations.}
\label{phase_diagram}
\end{figure*}

\color{black}
\subsection{Co-nonsolvency phase diagram}

To provide useful theoretical guidelines for co-nonsolvency in polymer brush systems, we constructed two co-nonsolvency phase diagrams using the better solvent fraction and the grafting density as control parameters (Fig.~\ref{phase_diagram}). Each of the three orientational phases is observed and they are separated by the dotted lines at $A = 0.0$ and $0.2$ as shown in the Fig.~\ref{phase_diagram}. Phases I, II, and III correspond to a parallel, isotropic, and perpendicular orientations of the polymer chains, respectively. The dotted lines indicate the boundary where the orientation of the polymer changes significantly. The regions with parallel or isotropic orientations (I, II) become narrower as the grafting density increases, eventually disappearing. At the highest grafting density, $A$ was about $1.0$, and the polymer is maintained perpendicular to the surface irrespective of the better solvent fraction. The narrow in-plane free zone arising from the high grafting density inhibits the orientation between polymers and interferes with solvent sharing with other polymers. Conversely, the isolation of grafted polymers from other polymers results in an isotropic arrangement at low grafting densities. As shown in Fig.~\ref{phase_diagram}, the co-nonsolvency occurs more effectively in the low grafting density environment where the polymer can change its orientation freely and interact with other polymers easily. However, note that this thickness space does not cover all specific co-nonsolvency systems, and the affinity between particles and the degree of polymerization can affect the structural response of the polymer.~\cite{Mukherji14,Mukherji15} Crucially, this diagram include the three characteristic regions: region (i), where the conformational response of chains correspond to that of the brush ($z_{\text c}/z_{\text s} \simeq R_{\text g,c}/R_{\text g,s}$); region (ii), Region where conformational response of chains correspond to that of the brush ($z_{\text c}/z_{\text s} > R_{\text g,c}/R_{\text g,s}$); region (iii), the non-response region, ($z_{\text c}/z_{\text s} \simeq 1$, where the subscript s and c mean that the value is observed at $x_{\text b}=0.0$ and $x_{\text b}=0.1$, respectively).

\section{Conclusion}

Understanding co-nonsolvency in realistic systems from a microscopic point of view is crucial for the application this fascinating phenomenon to diverse systems. Co-nonsolvency behavior results in a variety of response properties such as changes in the hydrodynamic radius, brush thickness, viscosity, frictional force, adhesion force, and growth homogeneity.~\cite{Scherzinger12, Sui11, Hao10, Yu16, Yu17, Pomorska16} Using a simplified coarse-grained model, we obtained results similar to experiment and successfully demonstrated that the subtle balance between the inter- and intramolecular bridging structures play a critical role in the co-nonsolvency behavior. Our findings reveal that controlling the grafting density enhances the structural response, and the response disappears above the critical grafting density. The surface-confined nature of the polymer brushes allows the regulation of the topologically excluded volume between polymers. We have also identified the novel scaling behavior $h \sim \sigma_{\text g}^{0.71}$ in the co-nonsolvent. We uncovered three orientational phases that are remarkably separated by two parameters: the solvent and grafting density. Overall, our simulation study provides microscopic insights into co-nonsolvency behavior and useful experimental guidelines for structurally adjusting the polymer orientations at a microscopic level.

In particular, drug delivery is a promising application of co-nonsolvency systems, especially PNIPAM brushes because of their bio-compatibility.~\cite{Lima16}. Important issues for the design of ideal drug delivery systems include identifying how the drug can be delivered at the right time and the right concentration.~\cite{Fu10,Luo12} For optimum therapeutic responses, it is important to study the diffusive behavior of solvent and the polymer disentanglement when the co-solvent is introduced to the system,~\cite{Miller-Chou03} which can be investigated explicitly using molecular dynamics simulations. Thus, we plan to explore the dynamical properties of the brush in co-nonsolvent mixtures further, such as determining how fast the brush response upon the co-nonsolvent stimuli, how the grafted chain fluctuation is correlated with the solvent environment, and whether the dynamic heterogeneity occurs as the spatial heterogeneity does in the co-nonsolvency window. In addition, we expect the many-chain effects on the dynamic properties to be crucial, as in the structural study.

\newpage 

\begin{acknowledgement}
This work was supported by Samsung Science and Technology Foundation (grant number SSTF-BA1601-11).
\end{acknowledgement}

\section{Supporting Information Available}

$\bullet$ System description\\ 
$\bullet$ Maximum height of the brush\\
$\bullet$ Equilibrium bulk density as a function of the better solvent fraction\\
$\bullet$ Brush height as a function of chemical potential of the better solvent\\ 
$\bullet$ Density profile of free ends\\ 
$\bullet$ Density profiles of monomer and solvents at xb=0.3, 0.5, 0.7\\
$\bullet$ Parallel and perpendicular components of the radius of gyration\\ 
$\bullet$ Schematic description of the bond angle\\ $\bullet$ Unlooping probability \\
$\bullet$ Differential looping probability for the loop size\\
$\bullet$ Monomer coordination number of solvent.

\bibliography{brush}

\end{document}